\documentclass[usenatbib,usegraphicx]{mn2e}
\usepackage{graphicx,verbatim,amsmath,amssymb,amsfonts}
\usepackage{subfigure}
\usepackage{float}
\usepackage{color}
\usepackage{enumitem}

\def\apjl{ApJL}
\def\apjs{ApJS}
\def\aap{A\&A}

\def\ltsima{$\; \buildrel < \over \sim \;$} 
\def\lsim{\lower.5ex\hbox{\ltsima}}
\def\gtsima{$\; \buildrel > \over \sim\;$} 
\def\gsim{\lower.5ex\hbox{\gtsima}} 
\def\msun{\rm M_\odot}
\def\zsun{\rm Z_\odot}

\defcitealias{Capak15}{C15}
\defcitealias{pallottini:2017}{P17}

\title[ALMA suggests outflows in $z\sim 5.5$ galaxies] 
{ALMA suggests outflows in $z\sim 5.5$ galaxies}  
\author
[S. Gallerani, et al.]
{S. Gallerani$^{1\ast}$, A. Pallottini$^{1,2,3}$,  C. Feruglio$^{1}$, A. Ferrara$^{1}$, R. Maiolino$^{2,3}$, L. Vallini$^{4,5,6}$, \newauthor D. A. Riechers$^{7}$, R. Pavesi$^{7}$
\vspace{10pt}\\
$^1$ Scuola Normale Superiore, Piazza dei Cavalieri 7, 56126, Pisa, Italy \\ 
$^2$ Kavli Institute for Cosmology, University of Cambridge, Madingley Road, Cambridge CB3 0HA, United Kingdom \\
$^3$ Cavendish Laboratory, University of Cambridge, 19 J. J. Thomson Ave., Cambridge CB3 0HE, United Kingdom\\
$^4$ Nordita, KTH Royal Institute of Technology and Stockholm University, Roslagstullsbacken 23, SE-106 91 Stockholm, Sweden\\
$^5$ Istituto Nazionale di Astrofisica - Osservatorio Astronomico di Bologna, via Ranzani 1, I-40127 Bologna, Italy\\
$^6$ Dipartimento di Fisica e Astronomia, Universit\'a di Bologna, viale Berti Pichat 6/2, 40127 Bologna, Italy\\ 
$^7$Department of Astronomy, Cornell University, 220 Space Sciences Building, Ithaca, NY 14853, USA\\
$^\ast$To whom correspondence should be addressed; E-mail: simona.gallerani@sns.it\\
}
\date{\today} 
\pagerange{\pageref{firstpage}--\pageref{lastpage}} \pubyear{2016} 
\begin{document} 
\maketitle 
\label{firstpage} 
\begin{abstract} 
We present the first attempt to detect outflows from galaxies approaching the Epoch of Reionization (EoR) using a sample of 9 star-forming ($\rm SFR=31\pm 20~M_{\odot}~yr^{-1}$) $z\sim 5.5$ galaxies for which the [CII]158$\mu$m line has been previously obtained with ALMA. We first fit each line with a Gaussian function and compute the residuals by subtracting the best fitting model from the data. We combine the residuals of all sample galaxies and find that the total signal is characterised by a flux excess of $\sim 0.5$ mJy extended over $\sim 1000$ km~s$^{-1}$. Although we cannot exclude that part of this signal is due to emission from faint satellite galaxies, we show that the most probable explanation for the detected flux excess is the presence of broad wings in the [CII] lines, signatures of starburst-driven outflows. We infer an average outflow rate of $\rm \dot{M}=54\pm23~ M_{\odot}~yr^{-1}$, providing a loading factor $\eta=\rm \dot{M}/SFR=1.7\pm1.3$ in agreement with observed local starbursts. Our interpretation is consistent with outcomes from zoomed hydro-simulations of {\it Dahlia}, a $z\sim 6$ galaxy ($\rm SFR\sim 100~\rm M_{\odot}~yr^{-1}$) whose feedback-regulated star formation results into an outflow rate $\rm \dot{M}\sim 30~ M_{\odot}~yr^{-1}$. The quality of the ALMA data is not sufficient for a detailed analysis of the [CII] line profile in individual galaxies. Nevertheless, our results suggest that starburst-driven outflows are in place in the EoR and provide useful indications for future ALMA campaigns. Deeper observations of the [CII] line in this sample are required to better characterise feedback at high-$z$ and to understand the role of outflows in shaping early galaxy formation.
\end{abstract} 
\begin{keywords} 
galaxies: ISM - galaxies: evolution - galaxies: high-redshift
\end{keywords} 
\section{Introduction} 

\begin{figure*}
\vspace{1.5cm}
\includegraphics[width=0.99\textwidth]{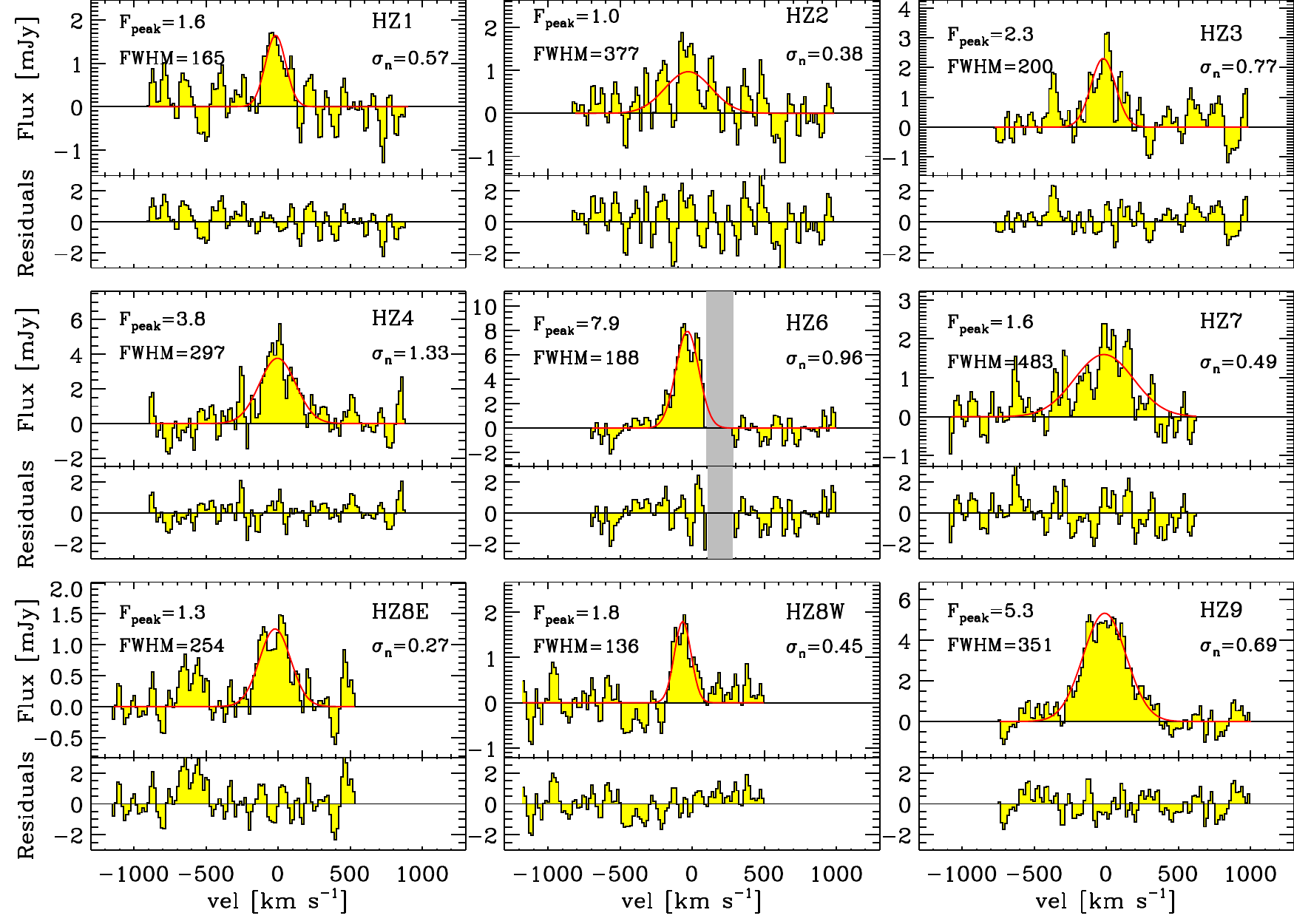}
\caption{
[CII] emission lines observed with ALMA by C15 (yellow shaded regions). The red solid lines represent the single Gaussian fit to the data. For each galaxy, the best fit parameters ($\rm FWHM$, $F_{\rm peak}$) and the noise ($\sigma_n$) are reported; the normalized residuals are shown in the bottom panels by the yellow shaded regions; the gray shaded region in HZ6 represents a frequency range in which the atmospheric transmission shows an enhanced dip in the case of $pwv=2$~mm.\label{fig1}}
\end{figure*}

\begin{figure*}
\includegraphics[scale=0.51]{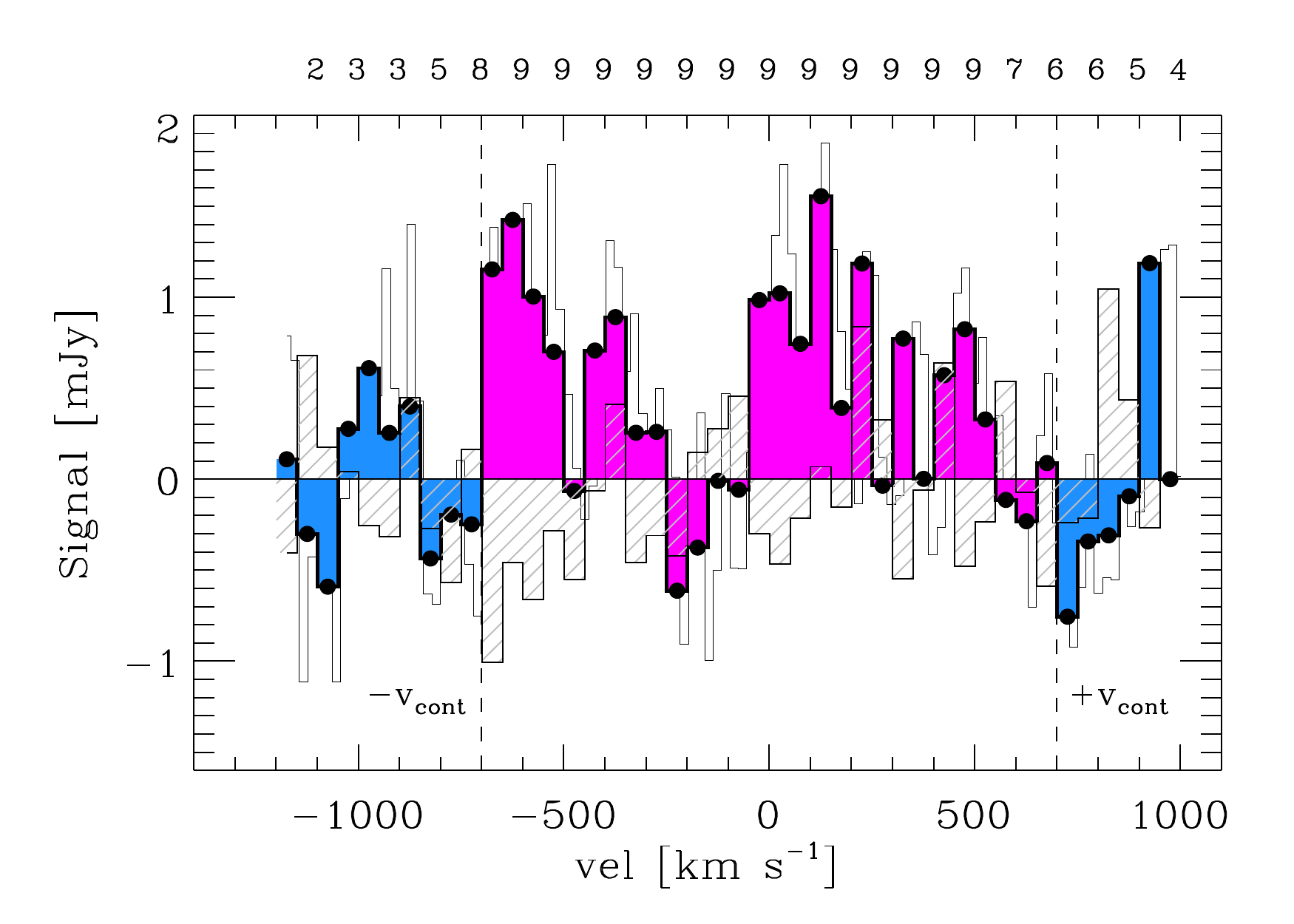}
\includegraphics[scale=0.48]{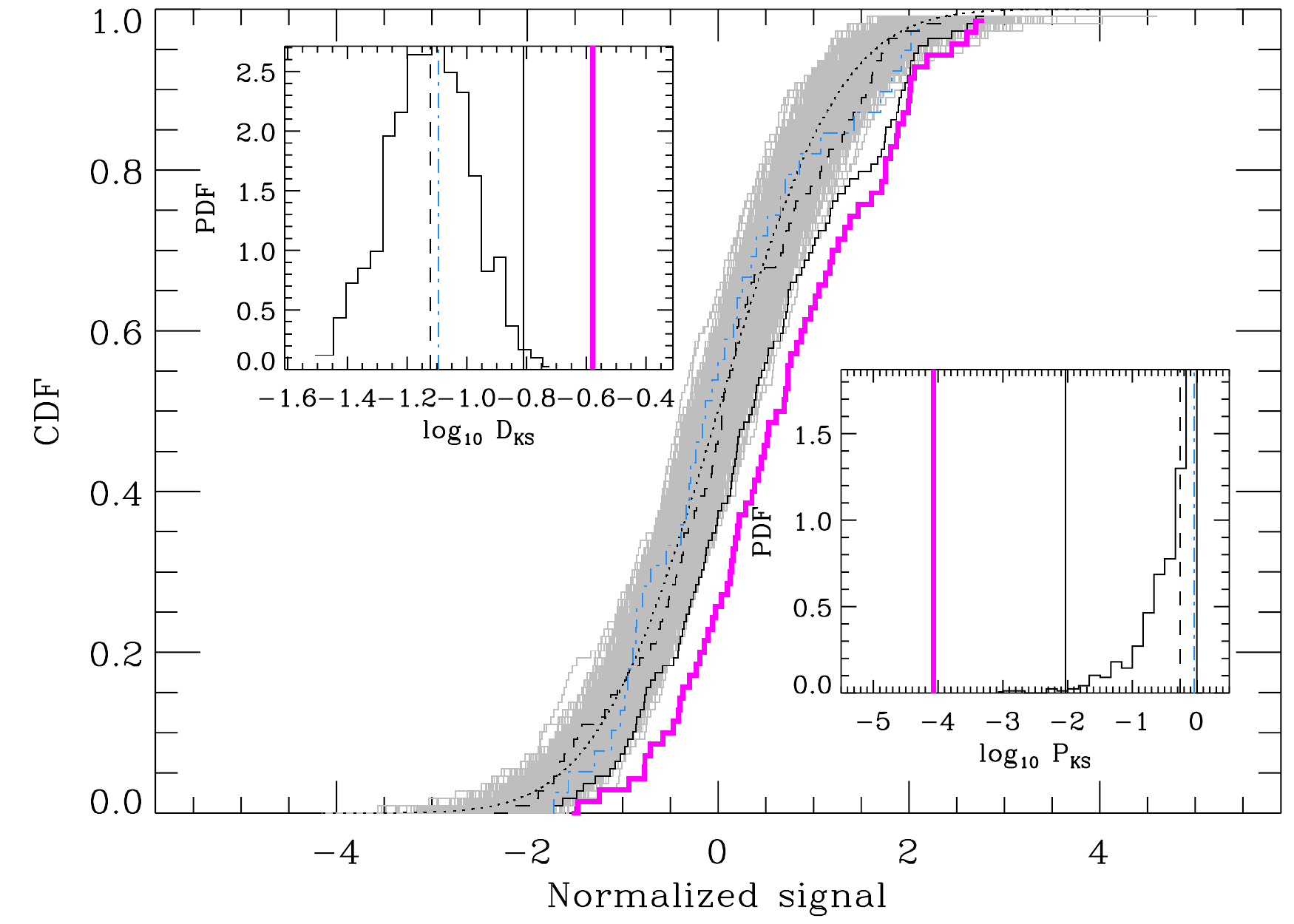}
\caption{{\bf Left panel:} The coloured histogram shows $R^{\rm tot}\times \langle\sigma_n\rangle$, i.e. the combined normalized residuals, multiplied for the mean value of the noise of all galaxies $\langle\sigma_n\rangle=0.7~\rm mJy$. The magenta region refers to pixels characterized by $\vert v \vert < v_{\rm cont}$, where $v_{\rm cont}=700~\rm km~s^{-1}$, while the blue region represents the region used for the noise determination, namely $\vert v \vert > v_{\rm cont}$. The gray hatched histogram represents $G^{\rm tot}$, i.e. the combined standard normal deviates (see footnote \textsuperscript{\ref{SND}}). Both signals have been rebinned to 50 $\rm km~s^{-1}$. For each spectral bin we report, on the top of the spectrum, the number of galaxies that contribute to the corresponding flux. The thin solid black line shows $R^{\rm tot}\times \langle\sigma_n\rangle$ at a resolution of 20~$\rm km~s^{-1}$. {\bf Right panel:} Cumulative distribution functions (CDFs) of $R^{\rm tot}$ (solid black line) and $G^{\rm tot}$ (dashed black line), rebinned to a spectral resolution of 20~$\rm km~s^{-1}$. The solid thick magenta (dot-dashed blue) line shows the CDF of $R^{\rm tot}$ computed in the velocity range $\vert v \vert < v_{\rm cont}$ ($\vert v \vert > v_{\rm cont}$). The shaded gray region represents the CDFs of 500 {\it standard normal deviates}, while the dotted line denotes the error function. See the main text for a description of the insets.
\label{fig2}
}
\end{figure*}

Massive stars profoundly affect the interstellar medium (ISM) of galaxies by injecting energy and momentum in the gas. This occurs via
radiation pressure, stellar winds, photoionization, and supernovae \citep{Dekel86, MacLow99, Murray11}. As a result the gas is heated and ionized, and it becomes turbulent. Moreover, a substantial fraction of the ISM can be cast into the halo \citep{Hopkins12,pallottini:2014a,liang:2016} and the intergalactic medium (IGM), thus enriching these components with freshly produced heavy elements \citep{Oppenheimer06, Pallottini14b}. Outflows also play a key role in the life-cycle of galaxies, as they control star formation by regulating the amount of gas available to form stars. On cosmological scales, outflows are thought to shape the faint-end of the galaxy luminosity function up to the highest redshifts at which galaxies can be observed \citep{Bouwens14, Dunlop13}. This complex network of physical processes is collectively known as ``feedback''.  

Theoretically, modeling feedback represents a formidable challenge. This is because its ab-initio implementation in cosmological simulations is hampered by the range of scales \citep[from Mpc to sub-pc;][]{Agertz13} involved in the problem, and also by the complexity of the physical network. Nevertheless, some simple scalings with global properties of galaxies have provided at least a phenomenological link to observations \citep{Dave12, Dayal14}. 

Observations provide crucial insights and guidance into outflow physics and driving. The most studied local starburst galaxy, M82, shows a prominent biconical, multiphase outflow. X-ray observations suggest the coexistence of hot ($\sim 10^7~\rm K$) gas together with a warm ($\sim 10^3~\rm K$) H$\alpha$-emitting phase \citep{Lehnert99} in which cold ($\sim 10~{\rm K}$), molecular clumps are embedded, as revealed by CO observations \citep{Walter02}. Evidence of outflows from local dwarf galaxies are also abundant (e.g. \citealt{Martin02}; see also review by \citealt{Veilleux05}). Finally, very fast (up to 1000-2000 km/s) AGN-driven outflows of molecular gas, extending on kpc scales, have been found in a dozen dusty star forming galaxies \citep{Veilleux13, Cicone14, Feruglio15}. 

At higher redshifts, detecting outflows becomes much more difficult. However, thanks mostly to absorption line spectroscopy, the presence of powerful outflows in galaxies close to the peak of the cosmic star formation ($z\approx 2-3$),  has been firmly assessed \citep{Shapley03,Steidel10,kacprzak:2014,rubin:2014,heckman:2015,schroetter:2015}. Pushing observations into the Epoch of Reionization (EoR, $z > 6$) is clearly the next frontier. This would be particularly important for several reasons. First, quasar absorption line experiments have shown that the IGM is already substantially enriched by the end of the EoR \citep{Dodorico13}, thus bringing a strong argument in favor of widespread galactic outflow activity. Second, a consensus exists that reionization has been primarily driven by ionizing photons produced by low-mass galaxies \citep{Choudhury07, choudhury:2008, Finkelstein12, Mitra12, Robertson15, Mitra15, Bouwens15} which are the most sensitive to feedback effects. Thus, outflows have likely regulated the reionization process both by modulating star formation, and the escape of LyC photons through the resulting hot cavities. 

Encouraged by the success in detecting outflows from high-$z$ quasars using sub-mm lines \citep{Maiolino12, Cicone15}, in this paper we present the first attempt to use ALMA data to detect outflows from the analysis of the [CII] 158$\mu$m line emitted by galaxies in the EoR. 

\section{ALMA observations}
We consider a sample of high-$z$ ($5.2\lesssim z\lesssim 5.7$) galaxies for which spectra of the [CII]158$\mu$m line have been obtained with ALMA at an angular resolution of\footnote{In this work we assume a $\Lambda$CDM model with cosmological parameters compatible with \emph{Planck} results: $\Omega_{\Lambda}= 0.692$, $\Omega_{m}= 0.308$, $\Omega_{b}= 0.0481$, Hubble constant $\rm H_0=100\,{\rm h}\,{\rm km}\,{\rm s}^{-1}\,{\rm Mpc}^{-1}$ with ${\rm h}=0.678$, spectral index $n=0.967$, $\sigma_{8}=0.826$ \citep[][]{planck:2014}. At $z\sim 5.5$, an angular resolution of $\sim$0.6'' corresponds to $\sim 3.7$~kpc.}  $\sim$0.6'' by \citet[][\citetalias{Capak15} hereafter]{Capak15}. We consider HZ8 and its ``companion'' galaxy HZ8W as two distinct sources. We do not consider the quasar HZ5, and the galaxy HZ10 since its [CII] emission line is located at the edge of the observed spectrum. The final sample consists of 9 star forming galaxies ($5\lesssim \rm SFR\lesssim 70~M_{\odot}~yr^{-1}$) characterized by a stellar masses $\sim 10^{10}~\rm M_{\odot}$.

The Common Astronomy Software Application (CASA) was used for data reduction and analysis. The continuum subtraction was done by creating a continuum image from the three line-free $\sim$2 GHz base-bands, and then subtracting a model thereof in the visibility plane. A linear subtraction (i.e. 0$^{th}$ order polynomial) was performed across the bandpass. We underline that the galaxies of the \citetalias{Capak15} sample exhibit a systematically low dust content: only three out of the nine sources considered (namely HZ4, HZ6 and HZ9) have been detected in continuum emission with fluxes $\sim$0.1-0.5 mJy. A Briggs flexible weighting scheme (Robust = 1) was adopted\footnote{In the case of HZ6, we have applied a natural weighting scheme (Robust = 2) to the ALMA data. We have also tried to fit the [CII] emission line extracted from a no continuum subtracted cube with a ``Gaussian + continuum" function. We have checked that different weighting schemes and/or continuum subtraction techniques do not affect the main result of this work.}. For what concerns the bandpass calibration, the shape in each band was measured on a calibrator, to divide out the instrumental response\footnote{The sources J0538-4405, J0522-3627, J1037-2934 were used for the bandpass calibration. From the ALMA calibrator source catalogue (https://almascience.eso.org/alma-data/calibrator-catalogue) it can be seen that these sources show flux variations of 50-65\% on a timescale of 6 years (i.e. $<$ 0.03\% per day). Thus, uncertainties related to calibrators variability are expected to be negligible. Moreover, we have checked that, within 1$\sigma$ flux density uncertainties, the spectra of the bandpass calibrators are flat. Ganymede, Pallas, and Callisto were used for the flux calibration (providing an accuracy of $\sim$5\%); J1058+0133 and J1008+0621 for the phase calibration.}. We have removed from the observed spectra those pixels where the atmospheric transmission shows enhanced dips at the relevant frequencies, assuming $pwv=2$~mm for the precipitable water vapor. This is a conservative assumption since the actual values for these observations were $pwv=0.2-1.9$~mm (mostly $<1$mm). Pixels removal has been necessary only for HZ6 (see the gray shaded region in Fig. \ref{fig1}).

\section{Method}\label{method}
Although [CII] emission is clearly detected ($\geq 6\sigma$) in all the galaxies considered, the quality of their spectra is not sufficient for a detailed analysis of the [CII] line profile in individual galaxies. To overcome this difficulty, we adopt an inferential statistical approach. First of all, we start from the {\it null hypothesis} that the [CII] lines of our sample are well described by a Gaussian profile.
\begin{table}
\caption{Results for the Kolmogorov-Smirnov (KS) and Anderson-Darling (AD) tests. \label{tab_AD_and_KS_test}}
\centering 
\begin{tabular}{l c c c} 
\hline\hline
& KS & KS & AD \\ 
& probability & statistics & statistics\\ 
\hline 
$G^{\rm tot}$& 0.55&0.07 & 0.01\\ 
$R^{\rm tot}$& $4\times 10^{-3}$&0.17& 0.12\\
$R^{\rm tot}$ ($\vert v \vert < v_{\rm cont}$)&$3\times 10^{-5}$&0.28&0.31\\
$R^{\rm tot}$ ($\vert v \vert > v_{\rm cont}$)& 0.90&0.08& 0.02\\
\hline
\end{tabular}
\end{table} 

For each galaxy in the \citetalias{Capak15} sample, we fit the [CII] emission line ($F_{\rm obs}$) with a Gaussian function ($F_{\rm mod}$), whose free parameters are the full width at half maximum, $\rm FWHM$, the peak flux, $F_{\rm peak}$, and the center velocity, $v_0$. Moreover, we quantify the noise ($\sigma_n$) of the spectrum by computing the standard deviation of the observed flux in pixels characterized by a velocity $\vert v \vert > v_{\rm cont}$, where\footnote{The value of $v_{\rm cont}$ has been chosen in order to minimize the contamination by broad wings. However, its specific value does not affect the results of our study.} $v_{\rm cont}=700~\rm km s^{-1}$. We find $0.3<\sigma_n/[\rm mJy]<1.3$ with a mean value $\langle\sigma_n\rangle=0.7~\rm mJy$. Finally, we compute the residuals by subtracting the best fitting model from the observed spectrum, and normalize them to the noise, $R=(F_{\rm obs}-F_{\rm mod}){\sigma_n}^{-1}$.  As a sanity check of the method, for each galaxy, we also compute a {\it standard normal deviate}\footnote{A {\it standard normal deviate} is a random number extracted by a Gaussian distribution having mean equal to zero and standard deviation equal to one. \label{SND}}, hereafter called $G$. 

In Fig. \ref{fig1} we show the results from the above procedure for each galaxy in the sample. In each panel, we plot the observed spectrum (yellow shaded region), the Gaussian best fit (red line) and the normalized residual (yellow shaded region in the sub-panel). For each galaxy, we report in the figure the best fit values of the peak flux $F_{\rm peak}$, the FWHM of the [CII] line, and $\sigma_n$.\\
We thus combine both $R$ and $G$ of all galaxies into single arrays: 
\begin{equation}
R^{\rm tot}(v)=\frac{1}{n_{\rm gal}}\Sigma_{\rm j=1}^{n_{\rm gal}} R^j(v); G^{\rm tot}(v)=\frac{1}{n_{\rm gal}}\Sigma_{\rm j=1}^{n_{\rm gal}} G^j(v). 
\end{equation}
\begin{figure*}

\vspace{-5.5cm}
\includegraphics[scale=1.]{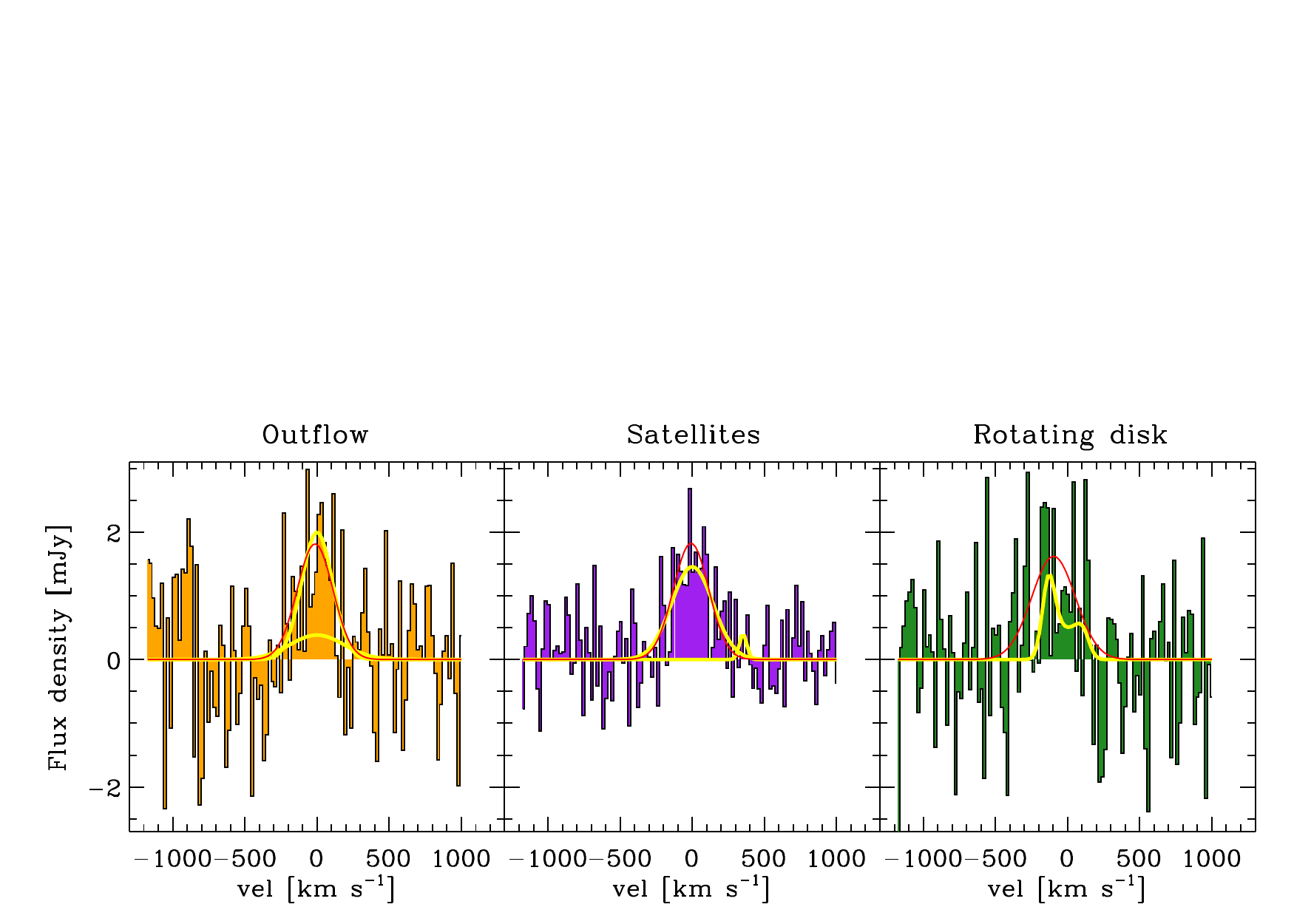}
\caption{Examples of synthetic spectra for the {\it Outflow} ({\bf left panel}), {\it Satellites} ({\bf middle panel}), and {\it Rotating disk} ({\bf right panel}) scenarios. The thick yellow lines represent the original synthetic spectra, while coloured shaded regions show the final spectra, after adding noise. The solid red lines represent the best-fit {\it single} Gaussian profile.}
\label{fig3}
\end{figure*}
The stacked signal, rebinned to 50 $\rm km~s^{-1}$, and multiplied for $\langle\sigma_n\rangle$ is shown in the left panel of Fig. \ref{fig2}, where we have differentiated pixels characterized by $\vert v \vert > v_{\rm cont}$ (blue shaded region), from those with $\vert v \vert < v_{\rm cont}$ (magenta shaded region). For each spectral bin, on top of the spectrum, we report the number of galaxies contributing to the corresponding flux. In the same figure, we plot $G^{\rm tot}\times \langle\sigma_n\rangle$ with an hatched gray region, rebinned to 50 $\rm km~s^{-1}$, and $R^{\rm tot}\times \langle\sigma_n\rangle$ with a solid black line at a resolution of 20 $\rm km~s^{-1}$. We find a flux excess of $\sim 0.5$ mJy extended over $\sim 1000$ km~s$^{-1}$.

We underline that the final stacked signal takes into account the differences between individual sources, since for each galaxy we subtracted the individual fit from the observed spectrum and normalized the residual to its own $\sigma_n$. Both in individual sources and in the stacked spectrum, the analysis takes into account the removal of a residual continuum term due to faint, individually undetected continuum or residuals in the subtraction.

If the observed spectra were completely determined by a single Gaussian [CII] emission line, the flux density of $R^{\rm tot}$ should consist simply of noise. In other words, both $R^{\rm tot}$ and $G^{\rm tot}$ should be described, by construction, by a {\it standard normal deviate}. We compute the cumulative distribution function (CDF) both of $R^{\rm tot}$ and $G^{\rm tot}$, at a resolution of 20~$\rm km~s^{-1}$, and we compare them with the error function:
\begin{equation}\label{erf}
erf(x)=\frac{2}{\sqrt\pi}\int_0^xe^{-t^2}dt,
\end{equation}
that describes the CDF of a {\it standard normal deviate}.

The resulting CDFs of $R^{\rm tot}$ (solid black line) and $G^{\rm tot}$ (dashed black line) are shown in the right panel of Fig. \ref{fig2}. The blue and magenta lines show the CDF of $R^{\rm tot}$ for $\vert v \vert > v_{\rm cont}$ and $\vert v \vert < v_{\rm cont}$, respectively, while the gray-shaded region is the result of the CDFs computed for 500 {\it standard normal deviates}. Finally, the dotted line represents the CDF described by eq. \ref{erf}. While the CDF of $R^{\rm tot}$ strongly differs from the $G^{\rm tot}$ CDF (particularly for $\vert v \vert < v_{\rm cont}$), the latter is, as expected, well within the gray-shaded regions. This sanity check ensures that the method adopted does not artificially introduces any deviation from a {\it standard normal deviate}.

To quantify the deviation of the CDFs from the error function {\it erf}, we apply the Kolmogorov-Smirnov (KS) test to our data: we compute i) the KS statistics, namely the maximum deviation ($D_{\rm KS}$) of the observed CDFs from the {\it erf} and the ii) the KS probability ($P_{\rm KS}$); small values of the KS probability imply that a CDF is significantly different from the {\it erf}. We apply the KS test to the CDF of $G^{\rm tot}$ and $R^{\rm tot}$ and we report the results in Tab. \ref{tab_AD_and_KS_test}. Moreover, we quantify the deviation of the CDFs of the random Gaussian deviates ${G_i}$ from the error function {\it erf}. In the bottom right and top left insets of Fig. \ref{fig2} (right panel), we show the probability distribution function (PDF) of $D_{\rm KS}^{\rm G_i}$ and $P_{\rm KS}^{\rm G_i}$, respectively, along with the results shown in Tab. \ref{tab_AD_and_KS_test}. The residual flux clearly exceeds that of a {\it standard normal deviates}. This result is also confirmed by applying the Anderson-Darling test to the data. According to this test, a sample is significantly different from a random gaussian deviate if the $AD$ statistics is larger then $0.05$ (see the third column in Tab. \ref{tab_AD_and_KS_test}).

\section{Interpreting the flux excess}\label{sec_fluxexc}

\begin{figure*}
\vspace{-5.cm}
\includegraphics[width=0.99\textwidth]{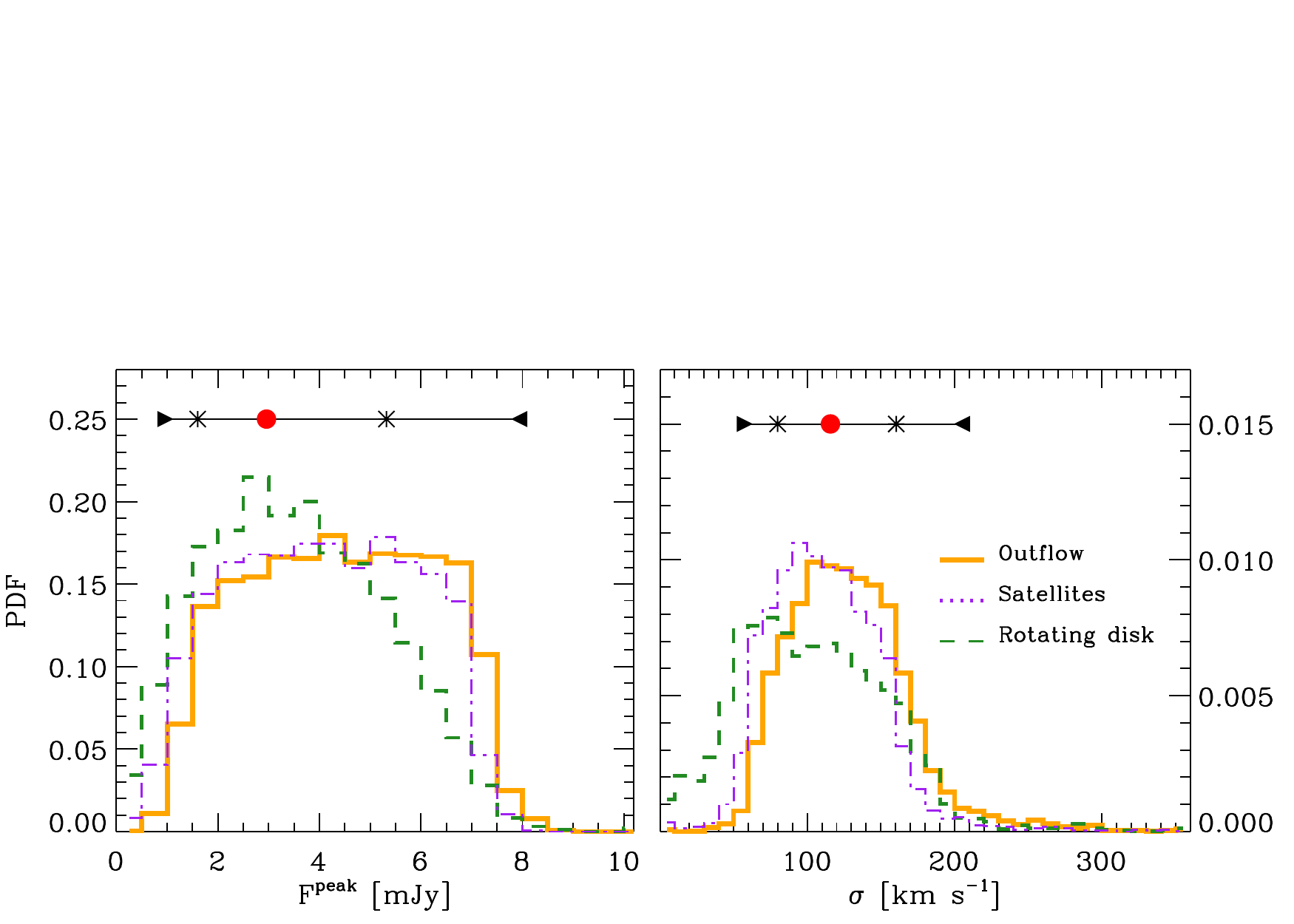}\\
\caption{Probability distribution functions (PDF) of the flux peak ($F^{\rm peak}$, {\bf left panel}) and width ($\sigma$, {\bf right panel}) of the {\it single} Gaussian component that provides the best-fit of the synthetic spectra for the models analysed. We plot the PDF for the {\it Outflow}, {\it Satellites} and {\it Rotating disk} scenarios with a solid orange, dotted green and dashed violet line, respectively. As a reference, for both $F^{\rm peak}$ and $\sigma$ we plot the min/1$^{\rm st}$ quartile/mean/3$^{\rm rd}$ quartile/max values obtained from the best-fit of the \citetalias{Capak15} sample with right triangle/asterisk/red circle/asterisk/left triangle, respectively.
\label{fig4}}
\end{figure*}
\begin{figure*}
\vspace{-5.5cm}
\includegraphics[scale=1]{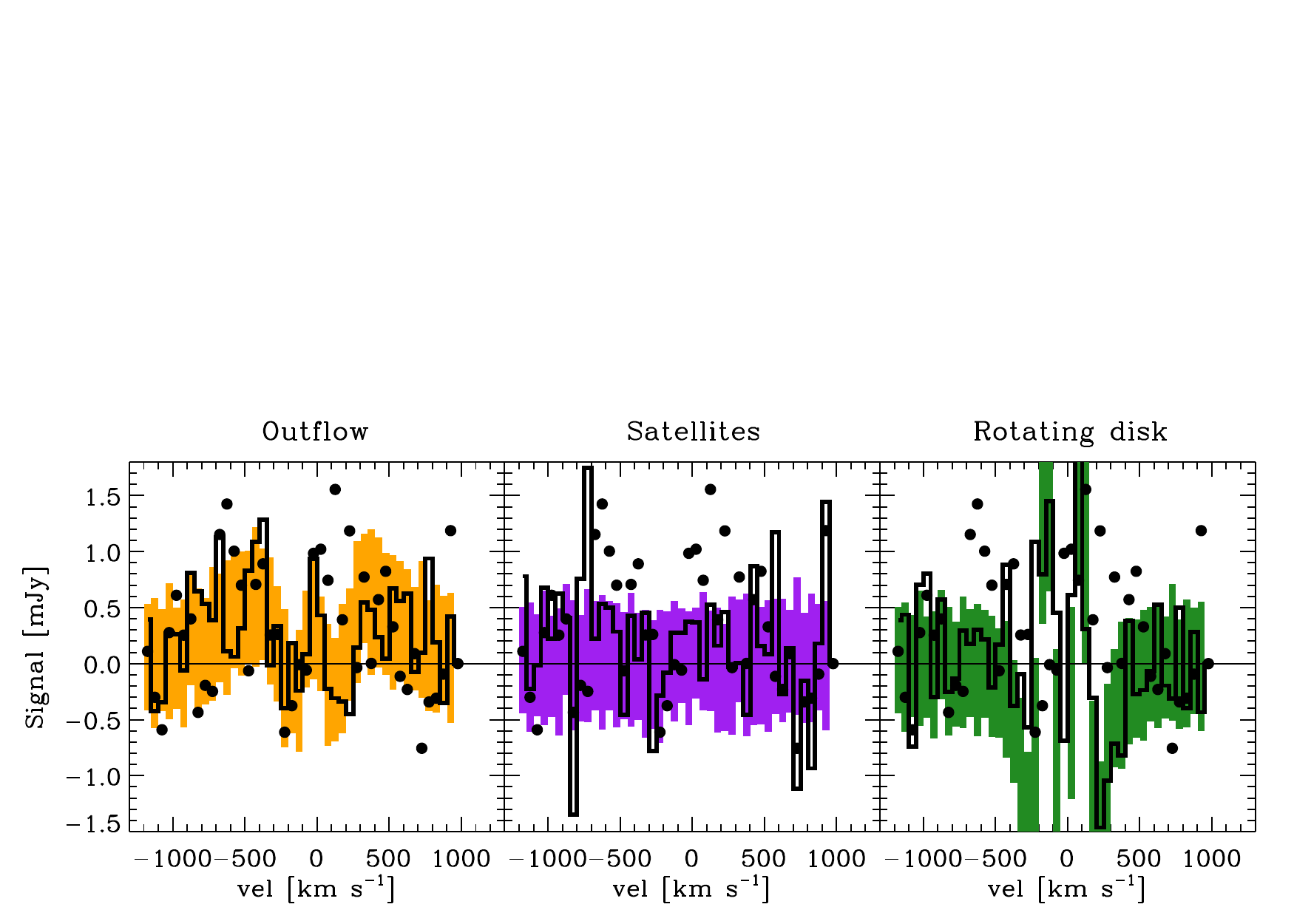}
\caption{Combined residuals for the synthetic samples (see left panel of Fig. \ref{fig2}). Filled black circles refer to the \citetalias{Capak15} sample. The black thick lines represent the realizations that provide the best agreement with data, while the coloured shaded regions quantify the 1$\sigma$ dispersion over the 500 realizations.}
\label{fig5}
\end{figure*}

\begin{table}
\caption{Parameter ranges for Monte Carlo simulations of synthetic profiles \label{tab_MC_parameters}} 
\centering 
\begin{tabular}{l c c c} 
\hline\hline
~ &{\it Outflow} & {\it Satellites} & {\it Rotating disk}\\ 
\hline 
$F_{\rm Peak}^{\rm narrow}~\rm[mJy]$&1--7&--&--\\ 
$\sigma^{\rm narrow}~\rm[km~s^{-1}]$&60--160&--&--\\
$F_{\rm Peak}^{\rm broad}~\rm[mJy]$&0.3--0.6&--&--\\ 
$\sigma^{\rm broad}~\rm[km~s^{-1}]$&100--500&--&--\\
\hline 
\hline
\# satellites &--&1&--\\ 
$F_{\rm Peak}^{\rm sat}~\rm[mJy]$&--&0.1--0.4&--\\ 
$\sigma^{\rm sat}~\rm[km~s^{-1}]$&--&20--50&--\\
\hline 
\hline
$F_{\rm Peak}^{\rm RD}~\rm[mJy]$&--&--&1--7\\ 
$v_{\rm c}~\rm[km~s^{-1}]$&--&--&50--200\\
$\sigma_{\rm gas}~\rm[km~s^{-1}]$&--&--&8--50\\
\hline 
\hline
\end{tabular}
\end{table} 

The reported flux excess leads us to reject the {\it null hypothesis} that a Gaussian model accurately describes the [CII] line profiles observed in the \citetalias{Capak15} sample. Thus, the next step is to look for {\it alternative hypothesis} that better describe the [CII] line shape of these high-$z$ galaxies. Several effects can explain the deviation of an emission line from a Gaussian profile:
\newpage
\begin{itemize}
\item if supernova-driven outflows are present as predicted \citep[e.g.][see also Sec. \ref{num_sim}]{pallottini:2017}, broad wings superposed to a narrower Gaussian core should feature the [CII] line profile;
\item a multi-peaked profile might result from the collective emission of satellite galaxies \citep[e.g.][]{vallini:2013,vallini:2015}; 
\item if a rotating disk is present, the line profile might take a double-horned profile \citep[e.g.][]{deblok:2014}.
\end{itemize}

To analyze these possibilities, we produce 3 sets of 500 Monte Carlo simulations of [CII] emission lines with different profiles, depending on the scenario considered (see Sec. \ref{dgprof}, \ref{satprof}, \ref{dhprof} below). To simulate the observed noise, we add to each synthetic spectrum a gaussian deviate with zero mean and $0.3<\sigma_n/[\rm mJy]<1.3$. For each scenario, we divide the full synthetic sample into sub-samples of 9 galaxies, and we apply to each sub-sample the same method described in Sec. \ref{method}: we first fit each synthetic profile with a single Gaussian (presented in Fig. \ref{fig4} in the next section); then, we normalize the residual to $\sigma_n$; finally, we stack the 9 simulated residuals into a single signal and we multiply it for $\langle\sigma_n\rangle$ (presented in Fig. \ref{fig5} in next Sec.).

\subsection{{\it Outflow}}\label{dgprof}
In this first scenario, synthetic [CII] emission lines are constituted by the sum of a narrow Gaussian (defined by its peak flux $F_{\rm peak}^{\rm narrow}$ and RMS $\sigma^{\rm narrow}$) plus a broad Gaussian profile ($F_{\rm peak}^{\rm broad}$; $\sigma^{\rm broad}$). We consider the parameter ranges shown in Tab. \ref{tab_MC_parameters}. We assume the same central velocity both for the narrow and broad component. We take $\sigma^{\rm broad}$ by numerical simulation results (see Sec. \ref{num_sim}). 

\begin{figure}
\includegraphics[scale=0.49]{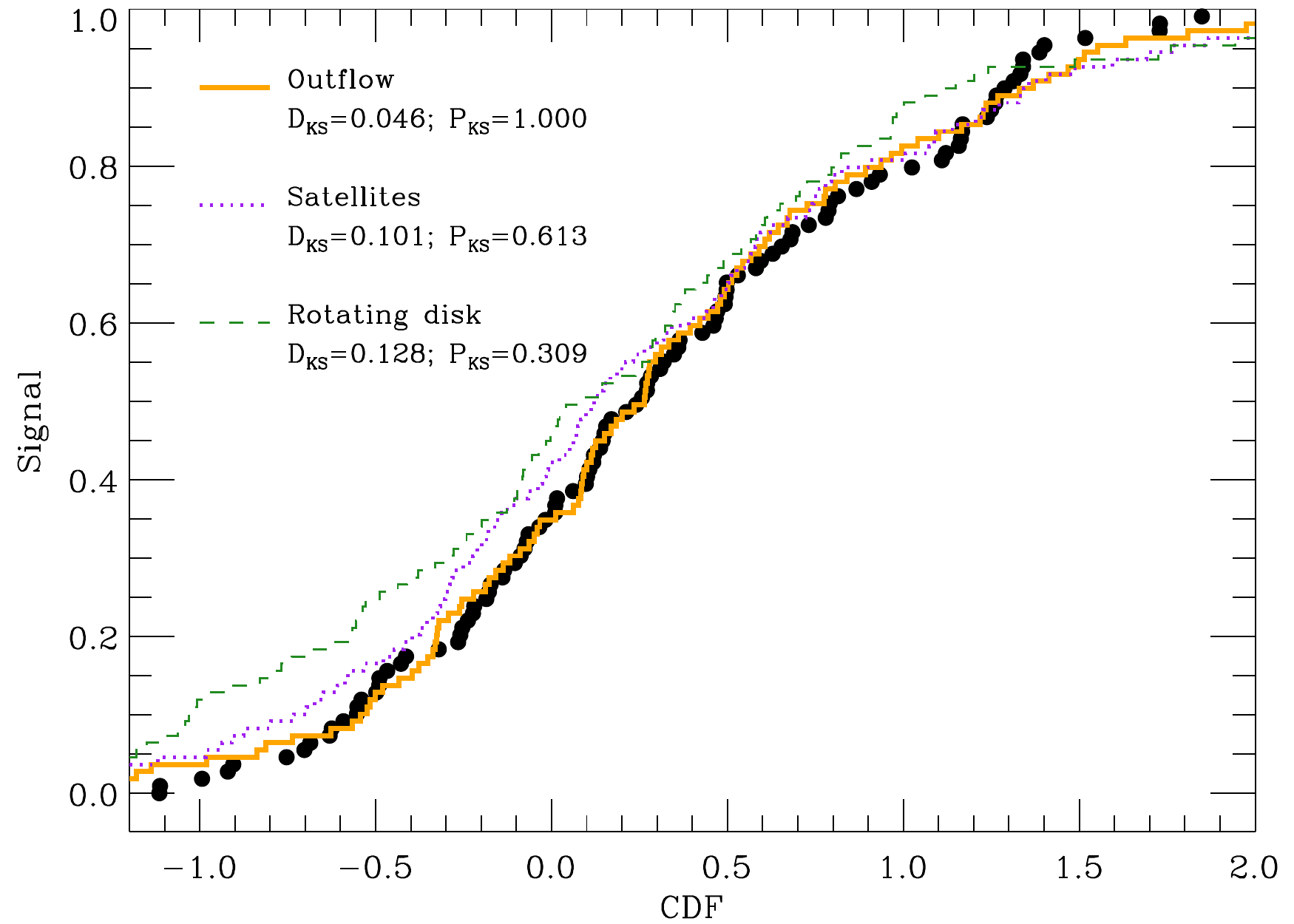}
\caption{CDF of the residuals for the synthetic samples (see right panel of Fig. \ref{fig2}). Filled black circles refer to the CDF of \citetalias{Capak15} sample. Coloured lines represent the realizations that provide the best agreement with data. We also report the results of the KS test (KS statistics, $D_{\rm KS}$, and KS probability, $P_{\rm KS}$) between the observed CDF and the simulated ones.
\label{fig6}
}
\end{figure}

\subsection{{\it Satellites}}\label{satprof}
To mimic the effect of satellites on the [CII] line profiles, we consider the sum of a Gaussian profile (defined by $F_{\rm peak}^{\rm narrow}$ and $\sigma^{\rm narrow}$) plus a number $N_{\rm sat}$ of Gaussians characterized by $F_{\rm peak}^{\rm sat}$ and $\sigma^{\rm sat}$ (see Tab. \ref{tab_MC_parameters}).\\
To constrain $\sigma^{\rm sat}$, we consider, as upper limit, the minimum value of $\sigma$ in the \citetalias{Capak15} sample ($\sigma=50~\rm km~s^{-1}$), and, as lower limit, the minimum value of $\sigma$ found in all the [CII] emission lines detected so far at high redshift \citep[$\sigma=20~\rm km~s^{-1}$;][]{pentericci:2016}.\\
For what concerns $F_{\rm peak}^{\rm sat}$, we use the $M_{\rm UV} - F_{\rm peak}$ relation presented in \citet[][see their eq. 2]{pallottini:2015}. The UV magnitude limit of the \citetalias{Capak15} sample ($M_{\rm UV}>-19.94$) converts into an upper limit for the [CII] emission line peak \citep[$F_{\rm peak}^{\rm sat}<0.4~\rm mJy$; see also][]{yue:2015,vallini:2015}. We note that this UV magnitude roughly corresponds to satellites with $SFR^{\rm sat}\sim 1~M_{\odot}~yr^{-1}$ and $M_*^{\rm sat}\sim 10^9 M_{\odot}$.\\
Finally, to compute the number of satellites that we expect to surround the galaxies of our sample, we adopt the halo occupation distribution model \citep[see eq. 14 by][]{yue:2013}. 
The \citetalias{Capak15} sample consists of $M_*\sim 10^{10}~M_{\odot}$ galaxies that are expected to be hosted in $M_{\rm DM}=6\times 10^{11}~\rm M_{\odot}$ dark matter halo. In such halo, we expect to find $\sim$4 satellites with $M_*^{\rm sat}\sim 10^9 M_{\odot}$. Given that these satellites can be located at a distance up to $r_{\rm vir}\sim 40$ kpc, we find that it is very unlikely to find any of them within 1-2 beams ($\sim$ 3-7~kpc) of the ALMA observations.\\
In our calculations, we use $F_{\rm peak}^{\rm sat}=0.1-0.4~\rm mJy$ and we assume that, for each of the \citetalias{Capak15} galaxies, the ALMA beam encloses a number of satellites $N_{\rm sat}=1$ (see Tab. \ref{tab_MC_parameters}). Thus, the contribution of satellites to the observed flux excess that we compute must be considered as a solid upper limit.
\subsection{{\it Rotating disk}}\label{dhprof}
To simulate the double-horned profile resulting from a rotating disk we refer to the work done by \citet[][in particular see eq. 1]{deblok:2014}. A double-horned profile is specified by its peak flux $F_{\rm peak}^{\rm RD}$, the disk circular velocity $v_{\rm c}$, and the gas velocity dispersion $\sigma_{\rm gas}$; the double-horned is given by the convolution of a gaussian ($F_{\rm peak}^{\rm RD}$,$\sigma_{\rm gas}$) with the velocity profile $\psi(v) = ( v^{2} - v_{\rm c}^{2} )^{-1/2} \Theta(|v_{\rm c}|-v) /\pi$, where $\Theta$ is the Heaviside function. We consider the parameter ranges shown in Tab. \ref{tab_MC_parameters}.

\subsection{Results}\label{results}

In Fig. \ref{fig3}, we show one example of synthetic emission line for each scenario described above ({\it Outflow} in the left, {\it Satellites} in the middle and {\it Rotating disk} in the right panel).

We model each synthetic spectrum with a single Gaussian profile: in Fig. \ref{fig4}, we show the PDF of the best-fit Gaussian peak (left panel) and $\sigma$ (right panel) (orange line for the {\it Outflow}, violet line for the {\it Satellites}, green line for the {\it Rotating disk} scenarios). As a reference, we plot the min, the 1$^{\rm st}$ quartile, the mean, the 3$^{\rm rd}$ quartile, and the maximum values resulting from the best-fit models of the \citetalias{Capak15} sample. The nice agreement between the observed values and the synthetic ones shows that the simulated lines have properties consistent with data.

We calculate the residuals using the best-fit Gaussians: in Fig. \ref{fig5} (left/middle/right panel) we compare the observed flux excess (filled circles) with the one we obtain by applying our method to the simulated samples ({\it Outflow}/{\it Satellites}/{\it Rotating disk}). The black thick lines represent the realizations that provide the best agreement with data, while the coloured shaded regions quantify the 1$\sigma$ dispersion over the 500 realizations. This figure shows that our method, in the case of the {\it Outflow} scenario, provides residuals that are in a qualitative better agreement with observations with respect to the {\it Satellites} and {\it Rotating disk} scenarios considered. 

To be more quantitative, in Fig. \ref{fig6}, we compare the observed CDF (filled circles) with the ones extracted from synthetic samples  (orange solid line for the {\it Outflow}, violet dotted line for the {\it Satellites}, green dashed line for the {\it Rotating disk} scenarios). We apply the KS test between the observed and simulated CDFs and we report the results in Fig. \ref{fig6}. We find that the distance from the observed CDF is minimized by the {\it Outflow} scenario ($D_{\rm KS}=0.05$). In this case, the probability that the simulated CDF reproduces the observed one is maximum ($P_{\rm KS}=1.0$). However, we cannot exclude that part of the flux excess we detect is due to emission from satellite galaxies. In fact, in the {\it Satellites} scenario we find $D_{\rm KS}=0.10$ and $P_{\rm KS}=0.6$. The {\it Rotating disk} scenario is instead more difficult to be reconciled with observations: $D_{\rm KS}=0.13$, $P_{\rm KS}=0.3$.

This analysis suggests that the observed flux excess can be ascribed to broad wings of the [CII] line tracing a starburst-driven outflow. In the realization that better describes the observed flux excess, the average value of the flux peak broad component is $\langle F_{\rm peak}^{\rm broad}\rangle=0.4\pm 0.1 \rm ~mJy$, while $\langle\sigma^{\rm broad}\rangle=360\pm 140 \rm ~km~s^{-1}$, for a total [CII] luminosity $L_{\rm [CII]}^{\rm broad}=2.9\pm 1.2 \times 10^8~L_{\odot}$.

As a final test, we have divided our sample in two sub-samples: S1 containing HZ1, HZ2, HZ3 and HZ7, HZ8 and HZ8W, i.e. galaxies with $SFR<50 \rm\, M_{\odot}\,yr^{-1}$, and S2 constituted by HZ4, HZ6 and HZ9, i.e. galaxies with $SFR\gtrsim 50\, \rm M_{\odot}\,yr^{-1}$. In both cases the deviation from a standard normal deviate is found, though the flux excess is stronger in the S2 sub-sample. This is expected as in the S2 sub-sample galaxies have more SNe available to drive outflows. However, the statistical significance of this result is limited by the small number of galaxies in the sub-samples.

\section{Numerical simulations}\label{num_sim}
\begin{figure*}
\centering
\includegraphics[width=0.45\textwidth]{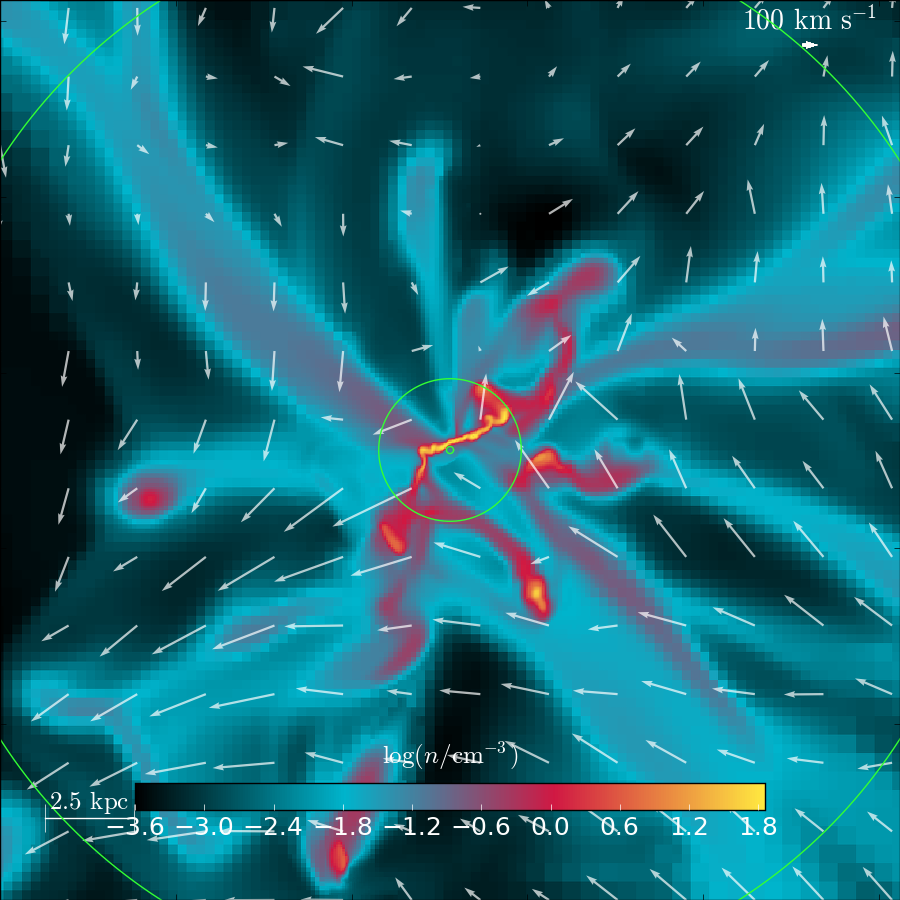}
\includegraphics[width=0.45\textwidth]{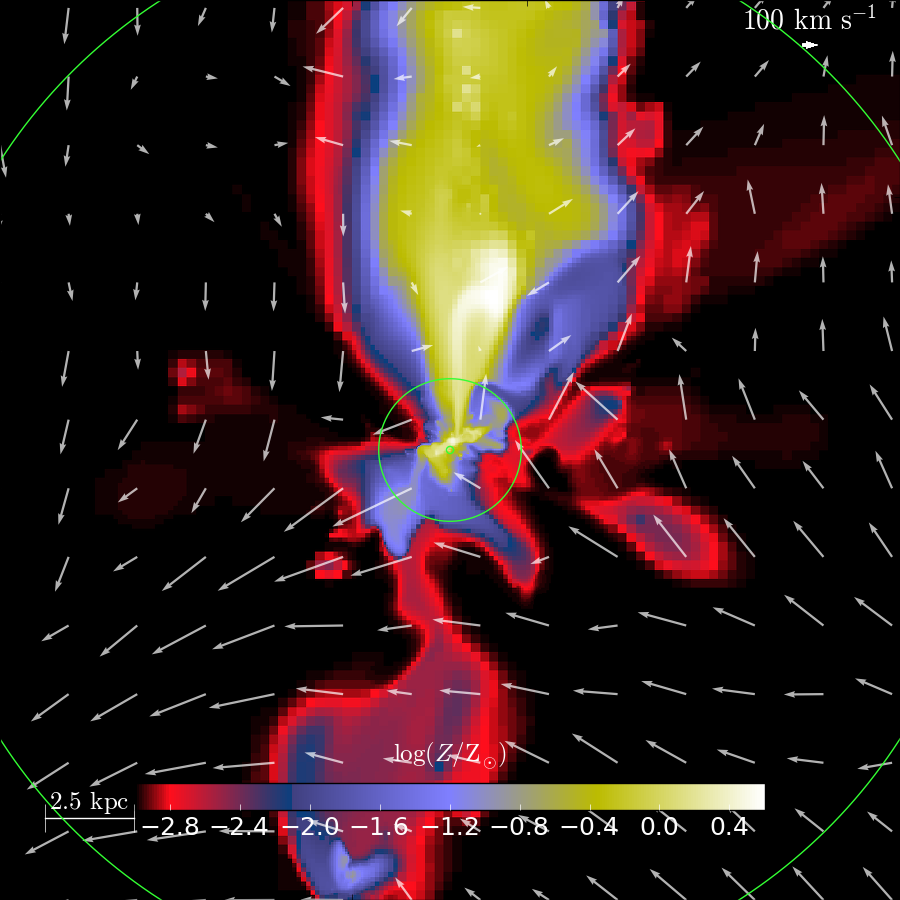}\\
\includegraphics[width=0.45\textwidth]{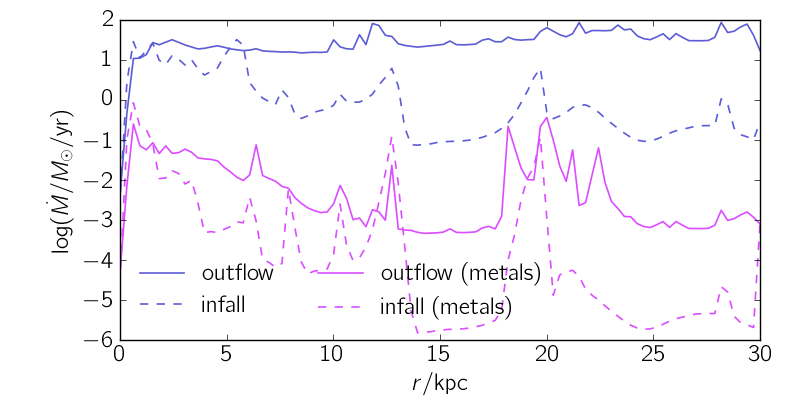}
\includegraphics[width=0.45\textwidth]{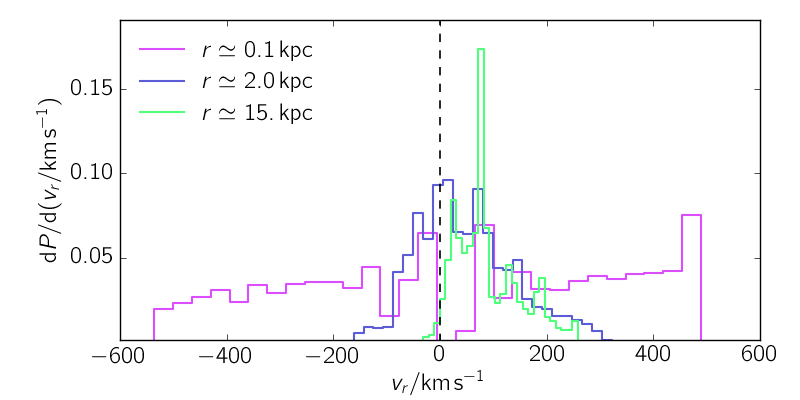}
\caption{
Slice of the density ($n/{\rm cm}^{-3}$, {\bf upper left} panel) and metallicity ($Z/\zsun$, {\bf upper right} panel) fields centered on the simulated galaxy \textit{Dahlia}. In both maps, the velocity field orthogonal to the line of sight is overplotted with white arrows and the spatial scale is indicated in the lower left corner. Green circles indicate a distance of $r/{\rm kpc}\simeq0.1,\,2,\,15~$ from the center. The {\bf lower left} panel shows the radially averaged outflow (solid line) and infall (dashed line) rate profiles for the gas (blue) and metals (magenta). The {\bf lower right} panel shows the PDF of the radial velocity ($v_r$) calculated for the gas at distance $r\simeq0.1~\rm kpc$ (magenta), $r\simeq2~\rm kpc$ (blue), $r\simeq15~\rm kpc$ (green).
\label{fig7}
}
\end{figure*}
We test our results against zoomed hydro-simulations of high redshift galaxies performed with the Adaptive Mesh Refinement (AMR) code RAMSES \citep{teyssier:2002}. These simulations are fully described in \citet[][\citetalias{pallottini:2017} hereafter]{pallottini:2017}; here, we summarize the aspects of the simulation that are relevant for this work.

Starting from cosmological initial conditions, we carry out a zoom-in simulation of a $z\sim 6$ dark matter (DM) halo of mass $\sim10^{11} \msun$ (virial radius of $\simeq 15~{\rm kpc}$). In the zoomed-in region, the gas mass resolution is $10^4 \msun$, and the AMR grid is refined to spatial scales $\simeq 30\,{\rm pc}$. We form stars from molecular hydrogen, following the model by \citet{krumholz:2009apj}. Stellar feedback includes supernovae, winds from massive stars and radiation pressure \citep[e.g.][]{Agertz13}. We model the thermal and turbulent energy content of the gas according to the prescriptions by \citet{agertz:2015apj}. We account for stellar energy inputs and yields that depend both on time and stellar populations \citep[e.g.][]{agora:2013arxiv}.

At $z\sim6$, the DM halo hosts a galaxy (named {\it Dahlia}) characterized by a stellar mass $M_*\sim 10^{10} \msun$ and a $\rm SFR\sim 100\, \msun {\rm yr}^{-1}$. In the top panels of Fig. \ref{fig7}, we show a slice of the density and metallicity fields (left and right, respectively) along with the velocity field orthogonal to the line of sight. Within a radial distance $r\lsim5~{\rm kpc}$ from {\it Dahlia}'s center, the gas has an average density of $n \simeq 10\,{\rm cm}^{-3}$, it is enriched to $Z\simeq 10^{-1} \zsun$, and rotates with a velocity $v_c\sim 100\,{\rm km}\,{\rm s}^{-1}$. At $5\lsim (r/{\rm kpc}) \lsim 15$, {\it Dahlia} is surrounded by a bubble of low density ($n \simeq 10^{-1}{\rm cm}^{-3}$), low metallicity ($Z\simeq 10^{-2} \zsun$) gas, outflowing at a speed $v\sim 100~\,{\rm km}\,{\rm s}^{-1}$. In the same region, dense ($n\simeq 1 {\rm cm}^{-3}$) and almost metal free $Z\simeq 10^{-3} \zsun$ filaments are infalling with a velocity $v\sim 100~\,{\rm km}\,{\rm s}^{-1}$. These filaments penetrate {\it Dahlia}'s circumgalactic medium at a distance of $\sim 7$ kpc (half of the virial radius), mixing with such component and becoming progressively more isotropic. The effect of the rapidly decreasing density and metallicity with the distance from the central of the galaxy is that negligible [CII] emission is expected on scales $\gtrsim$ 3 kpc \citep[see Fig. 10 and eq. 8 in][]{pallottini:2017}.

To better quantify our analysis, we compute the mass flow rate\footnote{The mass flow rate at a distance $r$ is defined as the material crossing a sphere of radius $r$: $\dot{M} = \int \rho {\bf v}\cdot d{\bf A}$, where $d{\bf A}$ is the surface area element, $\rho = \mu m_p n$ is the proton mass, and $v_{r}<0$ for infalling gas and $v_{r}>0$ for outflowing gas.} ($\dot{M}$) and the PDF of the radial velocity ($v_{r}$) at distances\footnote{To compute the radial velocity PDF we consider spherical shells with a thickness $\simeq 100\, {\rm pc}$ centered at a distance $r$ from {\it Dahlia}. This ensures that for each shell the PDF contains a minimum of $\sim$100 resolution elements.} $r\simeq0.1~\rm kpc$, $r\simeq2~\rm kpc$ and $r\simeq15~\rm kpc$. The results are shown in Fig. \ref{fig7}, in the bottom left and right panels, respectively.\\ At the center ($r\simeq0.1\,{\rm kpc}$), we find high radial velocities ($\sim 500\, {\rm km}~{\rm s}^{-1}$) for both the infall and outflow\footnote{We use these values of radial velocity to model the $\sigma^{\rm broad}$ in the {\it Outflow} scenario presented in Sec. \ref{dgprof}.}. As the infall fuels gas directly at the center, {\it Dahlia}'s SFR is mostly concentrated around $r\simeq0.1\,{\rm kpc}$. Via stellar feedback, this in turn causes high outflows velocities. Within $r\simeq 2~\rm kpc$ the mass inflow and the outflow rate are nearly equal ($\sim 30~\rm M_{\odot}~yr^{-1}$). Interestingly, the two rates are roughly equal also for metals. We interpret this as a result of efficient turbulent mixing between the two flows.\\ 
For $r\gsim2\,{\rm kpc}$ the velocity depends only mildly on radius, and since the density is approximately $\rho \propto r^{-2}$, the observed constancy of the outflow rate is mostly a consequence of a geometrical effect.\\ At $r\gsim5~\rm kpc$, the infall rate drops by a factor $\gsim 10$ below the outflow rate. This is caused by the different spatial structure of the two flows: while the incoming gas rains onto {\it Dahlia} along narrow filaments, the outflow is essentially spherically symmetric. Note that the outflow is $\sim10$ times more enriched than the infall, as most of the gas is essentially of pristine composition; thus, we expect the infall contribution to [CII] emission to be negligible with respect to the outflow. The ratio mildly declines with radius because of the progressive dilution of the outflow metal content with the IGM.\\
The presence of peaks at $r\sim 12~\rm kpc$ and $r\sim 20~\rm kpc$ is due to the presence of satellite galaxies. As projection effects are in place, we underline that it is quite unlikely to find such satellites in the $\sim 3\,\rm kpc$ beam centered on the main galaxy ({\it Dahlia} in this case) from which the [CII] spectra is calculated in \citetalias{Capak15}. From Fig. 10 by \citetalias{pallottini:2017}, we note that few ($\sim 3$) molecular clouds of mass $\sim10^6 M_\odot$ are present within $\lesssim 5~\rm kpc$ from the center of {\it Dahlia}; however, such clumps negligibly contribute to the [CII] emission since their typical luminosity is $L_{\rm CII}/L_{\odot} \simeq 10^5$ (see eq. 8 in \citetalias{pallottini:2017}), i.e. a factor $\sim 10^{3}$ less than the main galaxy.
\section{Conclusions}
We have analyzed the [CII] emission lines obtained with ALMA \citep{Capak15} in a sample of 9 star-forming ($5\lesssim \rm SFR\lesssim 70 ~M_{\odot}~yr^{-1}$) high-$z$ ($5.2\lesssim z\lesssim 5.7$) galaxies. We have fitted each [CII] emission line with a Gaussian function and we have computed the residual by subtracting the best fitting model from the data. By combining the residuals of all the galaxies, we have found a flux excess that can be ascribed to broad wings of the [CII] line tracing a starburst-driven outflow.
In order to get a rough estimate of the outflow rate $\dot{M}$, we use the following equation:
\begin{equation}
\dot{M}=v_{\rm outfl} \frac{M_{\rm outfl}}{R_{\rm outfl}},
\end{equation}
where $v_{\rm outfl}\sim 500~\rm km~s^{-1}$ is given by the largest velocity at which we observed the flux excess (see the left panel of Fig. \ref{fig2}), and $\rm M_{\rm outfl}=2.1\pm0.9\times 10^8~M_{\odot}$ is the total mass of the outflowing gas, obtained as in \citet[][see their eq. 1]{Maiolino12} considering the [CII] luminosity of the broad component estimated in Sec. \ref{results}. For what concerns $R_{\rm outfl}$, we assume that the spatial extent of the outflow is of the order of the radius of the deconvolved [CII] emission sizes, i.e. $\sim$ 1.9 kpc, a value consistent with the half-light radii of $z\sim 6$ LAEs \citep{taniguchi:2009,malhotra:2011,shibuya:2015}. Through this procedure, we infer an outflow rate of  $\rm \dot{M}=54\pm23~ M_{\odot}~yr^{-1}$. Given that the star formation rate of our sample is $\rm SFR=31\pm 20~M_{\odot}~yr^{-1}$, we estimate a loading factor $\eta=\dot{M}/SFR=1.7\pm1.3$, in nice agreement with the results found in local starbursts \citep[see top left panel of Fig. 6 by][]{heckman:2015}.

Moreover, \citet{heckman:2016} have recently analyzed a sample of low-$z$ ($0.2\lesssim z \lesssim 0.7$) galaxies characterized by a large range of SFRs ($10^{-2}\lesssim \rm SFR/[M_{\odot}~yr^{-1}]\lesssim 10^3$), and found a strong correlation between the specific SFR (sSFR=SFR/M$_*$) and $v_{\rm max}$ (see bottom left panel of their Fig. 1). From their scaling relation, we expect $v_{\rm max}\sim 500~ \rm km~s^{-1}$ for a $\rm sSFR\sim3\times 10^{-9}\rm yr^{-1}$, that is once more consistent with our findings. 

We further note that one galaxy of the sample used in this work, namely HZ1, also shows strong Ly$\alpha$ emission \citep[$L_{Ly\alpha}\sim 1.5\times10^8L_{\odot}$;][note that in their paper HZ1 is called N8bb-54-1862]{mallery:2012}. The Ly$\alpha$ line has an equivalent width $\rm EW_{Ly\alpha}\sim 5$~\AA~and results to be redshifted with respect to the [CII] line by $\sim$160 km~$\rm s^{-1}$, a value that is consistent with those reported for galaxies with similar UV luminosities and Ly$\alpha$ EWs (\citealt{erb:2014}; see also \citealt{willott:2015,pentericci:2016,carniani:2017}). Although the velocity offset between the Ly$\alpha$ line and the systemic redshift of galaxies could be ascribed to the presence of outflowing gas, it is important to underline that the Ly$\alpha$ line profile is affected by absorption from neutral hydrogen in the intergalactic medium and by absorption/scattering from dust in the interstellar medium \citep[e.g.][]{verhamme:2006}. Vice-versa, both processes do not affect the [CII] line profile. Thus, the origin of the velocity offset between the Ly$\alpha$ and [CII] lines is still argument of debate.

We test the results of our analysis against state-of-the-art hydrodynamical simulations of {\it Dahlia} \citep{pallottini:2017}, a star forming ($\rm SFR\sim 100~\rm M_{\odot}~yr^{-1}$) $z\sim 6$ galaxy. Star formation in the simulations is regulated by stellar feedback as radiation pressure and stellar winds from massive stars, and supernovae explosions. We find that, at a distance $r\lesssim 5~\rm kpc$ from the {\it Dahlia} center, the infall rate is balanced by outflowing gas with $\rm \dot{M}\sim 30~ M_{\odot}~yr^{-1}$. At larger distances, the outflow rate remains almost constant, while infalls are characterized by $\rm \dot{M}\sim 1~ M_{\odot}~yr^{-1}$. We underline that we can not resolve [CII] emission arising from scales smaller than the ALMA beam ($\sim$3-4 kpc). Thus the observed [CII] emission collects the contribution from all the material within this distance. Our simulations suggest that the flux excess resulting from our analysis at large velocities ($\sim$500 km~s$^{-1}$) mostly arises from gas located at small distance from the center of the galaxy ($\sim$0.1 kpc), while slower moving gas (still with velocities up to $\sim$100 km~s$^{-1}$) can be found at larger distances ($\gtrsim$~2 kpc).

The simulation outcomes support our interpretation of the ALMA data suggesting that starburst-driven outflows are in place in the EoR. However, we warn here that we are comparing an observed signal stacked from 9 galaxies with a single synthetic galaxy. For a more specific comparison, we urge to get deeper observations of the [CII] line in the galaxies of this sample to better characterize stellar feedback at high-$z$ and to understand its role in shaping early galaxy formation.

It would be interesting to apply the stacked analysis discussed in this work to other samples of galaxies and/or to extend the sample considered here. Other [CII] emitting galaxies ($\sim 14$) have been indeed discovered at high-$z$ in the last few years (e.g. \citealt{maiolino:2015,willott:2015,pentericci:2016,knudsen:2016,aravena:2016,bradac:2017}; see also Tab. 3 in \citealt{olsen:2017} for an updated list of [CII] observations in $z\gtrsim 5$ star forming galaxies). However, these observations span a large range in terms of redshift ($6<z<8$) and SFR (from few to almost thousands $\rm M_{\odot}yr^{-1}$). Thus, to date, the \citetalias{Capak15} sample represents the largest, homogeneous sample available for this kind of study.

\section*{Acknowledgements}
The authors are grateful to the anonymous referee for her/his insightful comments that have provided directions for additional work, improving the strength of our analysis. We thank F. Walter and A. Bolatto for useful discussions. This research was supported in part by the National Science Foundation under Grant No. NSF PHY11-25915. R.M. acknowledges ERC Advanced Grant 695671 ``QUENCH" and support by the Science and Technology Facilities Council (STFC). This paper makes use of the following ALMA data: ADS/JAO.ALMA\#2012.1.00523.S. ALMA is a partnership of ESO (representing its member states), NSF (USA) and NINS (Japan), together with NRC
(Canada), MOST and ASIAA (Taiwan), and KASI (Republic of Korea), in cooperation with the Republic of Chile. The Joint ALMA Observatory is operated by ESO, AUI/NRAO and NAOJ. The National Radio Astronomy Observatory is a facility of the National Science Foundation operated under cooperative agreement by Associated Universities, Inc.
\bibliographystyle{mn2e}
\bibliography{outflow_rev}
\bsp
\label{lastpage}
\end{document}